# Statistical models for RNA-seq data derived from a two-condition 48-replicate experiment

Marek Gierliński[1,2,†], Christian Cole[1,†], Pietà Schofield[1,2,†], Nicholas J. Schurch[1,†], Alexander Sherstnev[1], Vijender Singh[2], Nicola Wrobel[5], Karim Gharbi[5,6], Gordon Simpson[3], Tom Owen-Hughes[2], Mark Blaxter[5,6] and Geoffrey J. Barton[1,4,*]

[1]Division of Computational Biology; [2]Centre for Gene Regulation and Expression; [3]Division of Plant Sciences; [4]Biological Chemistry and Drug Discovery, College of Life Sciences, University of Dundee, Dow Street Dundee, DD1 5EH, UK. [5]Edinburgh Genomics, Ashworth Laboratories, University of Edinburgh, Edinburgh, UK. [6]Institute of Evolutionary Biology, Ashworth Laboratories, University of Edinburgh, UK. [†]These authors contributed equally to this work.

**ABSTRACT**

High-throughput RNA sequencing (RNA-seq) is now the standard method to determine differential gene expression. Identifying differentially expressed genes crucially depends on estimates of read count variability. These estimates are typically based on statistical models such as the negative binomial distribution, which is employed by the tools *edgeR*, *DESeq* and *cuffdiff*. Until now, the validity of these models has usually been tested on either low-replicate RNA-seq data or simulations. Here, a 48-replicate RNA-seq experiment in yeast was performed and data tested against theoretical models. The observed gene read counts were consistent with both log-normal and negative binomial distributions, while the mean-variance relation followed the line of constant dispersion parameter of ~0.01. The high-replicate data also allowed for strict quality control and screening of "bad" replicates, which can drastically affect the gene read-count distribution.

RNA-seq data have been submitted to ENA archive with project ID PRJEB5348.

**Contact:** g.j.barton@dundee.ac.uk

## 1 INTRODUCTION

High-throughput sequencing of RNA (RNA-seq, Nagalakshmi et al., 2010) estimates gene expression by counting the number of sequenced RNA fragments that map back to a given gene within a reference genome or transcriptome (Mortazavi et al., 2008). Differential gene expression (DGE) experiments compare this relative measure of transcriptional activity across several biologically interesting conditions to attempt to identify those genes that are fundamental to the difference between the conditions. This task is complicated by expression noise resulting from biological and technical variability, which introduces a level of uncertainty that has to be taken into account when the expression values from two or more conditions are compared. There are two major obstacles to identifying whether the observed difference in the expression of a gene between the two conditions is statistically significant, or whether it is consistent with arising through chance. Firstly, the observed read counts are a product of both the library size (total number of reads) and the fractional expression of the gene. The gene expression values thus need to be appropriately normalized before they can be meaningfully compared across conditions. Secondly, due to cost, time and workload constraints RNA-seq experiments typically consist of only a few replicates, making DGE genes particularly difficult to identify due to a lack of statistical power in the experiment (Hansen et al., 2011). These obstacles have motivated the development of numerous algorithms and computational tools to calculate DGE, each with their own underlying assumptions and their own approaches to data normalization and DGE detection. In particular, these tools commonly make assumptions about the form of the underlying of read-count distribution.

If sequenced RNA-seq reads originate randomly from the transcripts of expressed genes in a sample, the resulting read-count distribution would be multinomial, with parameters representing the proportions of reads mapping to individual genes. The observed read counts for an individual gene are then represented by a binomial random variable which, for a large total number of reads with only a small fraction of reads mapping to each gene, can be well approximated by a Poisson distribution. A key property of the Poisson distribution is that the variance is equal to the mean.

Although they have limited statistical power due to low numbers of replicates (Marioni et al., 2008, Robinson and Smyth, 2008, Rapaport et al., 2013), previous studies have shown that RNA-seq read-count data shows significant excess variance above that expected based on the Poisson model,. The excess variance observed in the data originates from biological processes, sample preparation, the sequencing protocol and/or the sequencing process itself. A natural over-dispersed alternative to the Poisson model is a negative binomial distribution, in which the variance is always greater than or equal to the mean.

The underlying read-count distribution for a gene is a fundamental property of RNA-seq data but without a large number of measurements it is not possible to identify the form of this distribution unambiguously. The limited number of replicates considered in previous studies means the true distribution of read counts for an individual gene is still unclear. Many DGE tools make strong assumptions about the form of this underlying distribution, including Poisson, negative-binomial and log-normal, which may impact on their ability to correctly identify significantly DE genes. In this study, the form of the underlying read count distribution is measured directly with a high-replicate, carefully controlled, RNA-seq experiment designed specifically for the purpose of testing the assumptions intrinsic to RNA-seq data. For the first time, these data allow various statistical models of read-count distribution to be tested directly against real RNA-seq data.

## 2 METHODS

### 2.1 Experiment design and data workflow

A controlled, 48 biological replicate, RNA-seq experiment in the model organism *Saccharomyces cerevisiae* was performed for wild type (WT) and a *snf2* knock-out mutant cell line (Δ*snf2*). Briefly, the extracted total RNA from each of the 96 replicates was enriched for polyadenylated RNA, quality checked, and had an appropriate amount of artificial ERCC spike-in transcripts added (Jiang et al., 2011, Loven et al., 2012) before undergoing the standard Illumina multiplexed TruSeq library preparation. The libraries were pooled and sequenced on 7 lanes of a single flow-cell on an Illumina HiSeq 2000 using a balanced block design (Auer and Doerge, 2010), resulting in a total of ~1 billion 50-bp single-end reads across the 96 samples. These reads were aligned to the Ensembl v68 (Flicek et al., 2011) release of the *S. cerevisiae* genome (modified to include the ERCC transcript sequences) with *TopHat2* (v2.0.5, Kim et al., 2013) The aligned reads were then aggregated with *htseq-count* (v0.5.3p9, Anders et al., 2014) using the Ensembl v68 *S. cerevisiae* genome annotation to give total gene read counts for 7,126 gene features. A full description of the experiment is provided in Schurch et al. (2015).

### 2.2 Notation

The following notation is used from this point on: $n_r = 48$ is the total number of replicates in each condition, $n_l = 7$ is the total number of lanes on the flowcell. The subscript $g$ denotes an individual annotated gene, $i$ and $j$ identify individual biological replicates within a given experimental condition, $(i, j = 1, \ldots, n_r)$ and $l$ is an individual lane on the flow cell ($l = 1, \ldots, n_l$). $x_{gil}$ denotes the number of read counts in gene $g$, replicate $i$ and lane $l$ and the read count for each biological replicate summed over all lanes is given by $x_{gi} = \sum_{l=1}^{n_l} x_{gil}$. Condition indices are skipped for clarity and the condition, for which the calculations are performed for, is specified in the text.

### 2.3 Normalization

RNA-seq experiments do not measure absolute gene expression; instead the number of sequenced reads mapped to a gene is proportional to its expression level, the sequencing depth and (in the case of short read sequencing) the length of the gene. Normalizing the data to compensate for these factors is a critical step in analysing RNA-seq expression data and it is essential that the normalization chosen is appropriate to the biological question being addressed. The simplest normalization scales the raw counts to the total number of sequenced reads in the replicate (total count normalization), however this is unlikely to be appropriate for a DGE experiment where one might reasonably anticipate that a small number of highly expressed genes are likely to change their expression level dramatically while the expression of the majority remains unchanged. Applying the total count normalization may result in many genes being identified as differentially expressed between conditions when they are, in fact, unchanged. Anders and Huber (2010) described a normalization strategy to address this based on the median expression across all genes, making it insensitive to highly expressed outliers. Other methods to tackle this problem are the trimmed mean of M-values (Robinson and Oshlack, 2010) and quantile normalization (Smyth, 2004). See Dillies et al. (2013) for a detailed review of these relevant normalization methods.

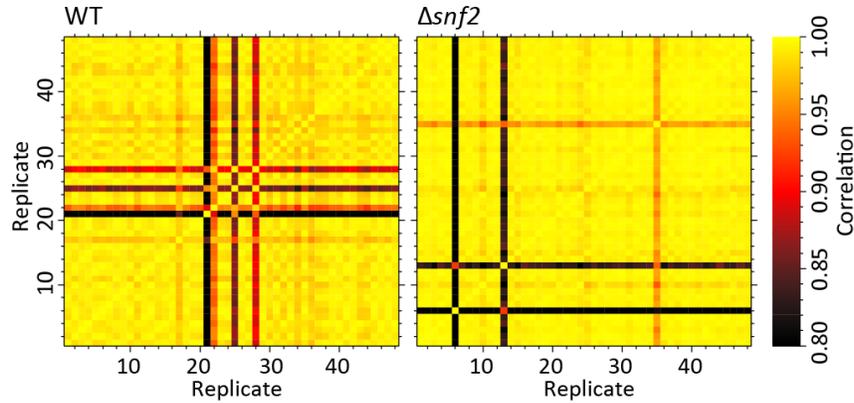

Fig. 1. Pearson's correlation coefficient between biological replicates in condition WT (left) and Δ*snf2* (right). Dark bands show "bad" replicates that poorly correlate with others.

These normalization approaches all result in counts that are no longer integer and this can lead to unpredictable results when testing them against discrete probability distributions (e.g., Poisson and negative binomial models). Accordingly, an alternative approach to normalization was applied, that while not suitable for DGE, is more appropriate for testing the underlying distribution of RNA-seq read counts for a gene. When examining the distribution of gene read counts across the biological replicates within a condition, the replicate with the smallest total read count was selected and then the other replicates were randomly sub-sampled to the same total read count. The same approach was applied to read counts across flow cell lanes within each biological replicate. The equal-count data are then mapped to the reference genome obtaining normalized, but discrete, mapped read counts.

The obvious disadvantage of this method is the loss of reads (~16% and 35% for lane and biological replicate data, respectively) and sensitivity. This approach is essentially a total count normalization that preserves the discrete nature of the data and so is unlikely to be any more appropriate for DGE analysis than standard total count normalization. However, under the null hypothesis, the down-sampled replicates within a condition are realizations of independent and identically distributed random variables, so it allows goodness-of-fits tests to be performed on raw gene counts with no need for any additional normalization. All the tests in this work are done within one biological condition (or one biological replicate, for lane data) and no DGE analysis is undertaken.

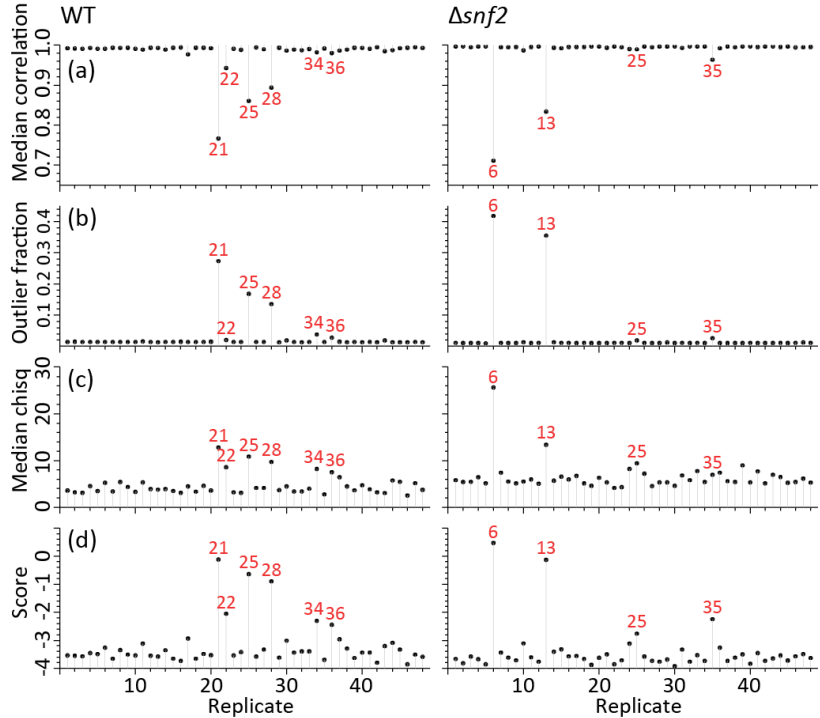

Fig. 2. Identifying bad RNA-seq replicates in the WT (left) and Δ*snf2* (right) data. The top three panels (a-c) show individual criteria for identifying "bad" replicates which are combined into a quality score (d) in order to identify "bad" replicates in each condition. The identified "bad" replicates are shown as numbered points in each panel. The individual criteria are (a) median correlation coefficient, $\tilde{r}_i$, for each replicate $i$ against all other replicates, (b) outlier fraction, $f_i$, calculated as a fraction of genes where the given replicate is more than 5 trimmed standard deviations from the trimmed mean and (c) median reduced chi-squared of pileup depth, $\tilde{\chi}_i^2$, as a measure of the non-uniformity of read distribution within genes (see also Fig. 3).

## 2.4 Quality control: identifying "bad" replicates

Occasionally errors occur in an experimental protocol and a sample generates improper or "bad" data. In this experiment, the availability of 48 replicates allowed three different criteria to be applied individually and in combination to identify potential "bad" replicates in addition to standard quality checks by FastQC.

### Replicate correlation

The differences between each pair of replicates, $i$ and $j$, is quantified by Pearson's correlation coefficient, $r_{ij}$, calculated across all genes with a non-zero count in at least one replicate (Fig. 1). The replicate's similarity to other replicates is captured by its median correlation, $\tilde{r}_i = \underset{j \neq i}{\text{median}}\, r_{ij}$ (Fig. 2a).

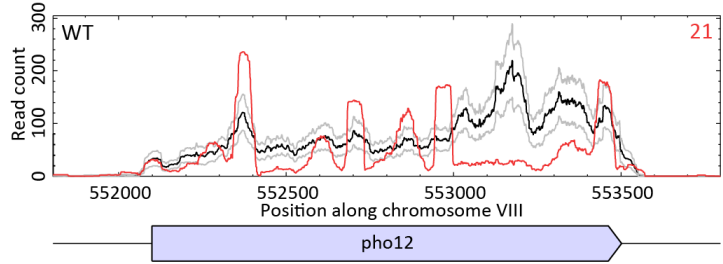

Fig. 3. Read depth profiles of YHR215W (*PHO12*). The black line indicates the mean read counts from all "clean" WT replicates for a given genomic position in the gene YHR215W (the set of "clean" replicates is defined in Section 3.1). The grey lines show the mean read depth plus/minus one standard deviation. The red line illustrates the read depth profile from a single example "bad" replicate (WT replicate 21). The block diagram at the bottom shows the simple gene structure of YHR215W.

## Outlier fraction

The poor correlations shown by some replicates are the result of a small proportion of genes with atypical read counts. These outliers can be identified by comparing each gene's expression in an individual replicate with the trimmed mean across all replicates. Specifically, the $n_t$ largest and smallest values are trimmed from the set of replicates for a gene before calculating the mean ($\bar{x}_{g;n_t}$) and standard deviation ($s_{g;n_t}$). Genes are then identified as outliers if $|x_{gi} - \bar{x}_{g;n_t}| > n_s s_{g;n_t}$, where $n_s$ is a constant. Fig. 2b shows the fraction $f_i$ of all genes identified as outliers for each replicate $i$, for $n_t = 3$ and $n_s = 5$. As expected, the anomalous replicates with high outlier fraction in Fig. 2b correspond well with the poorly correlating replicates in Fig. 2a. Increasing $n_t$ and/or decreasing $n_s$ enhances the outlier fraction in replicates already identified as anomalous; for example, reducing the standard deviation limit to $n_s = 3$ boosts the outlier fraction in these replicates by a factor ~2–3.

## Gene read depth profiles

The atypical total read counts of outlier genes are, in our case, the result of an atypical, strongly non-uniform, read depth gene profile. An example of the difference of the read depth profiles between clean and "bad" replicates is shown in Fig. 3 for the gene YHR215W. The distribution of reads from the example "bad" replicate (WT replicate 21, red line) is much less uniform than the mean read depth from the other "clean" replicates (black line) with distinct peaks in the distributions that are 50 bp long. These suspicious features are found universally in all outlier genes in "bad" replicates and can be also seen in some (but not all) non-outlier genes in these replicates. It is likely that the cause of this atypical read distribution is uneven priming during the PCR amplification step of the library preparation, prior to the sequencing. The level of gene non-uniformity in each replicate can be quantified with a reduced chi-squared statistic for each gene in each replicate, defined as

$$\chi^2_{gi} = \frac{1}{l_g - 1} \sum_{p=1}^{l_{gi}} \frac{(x_i - \bar{c}_{gi})^2}{\bar{c}_{gi}}, \tag{1}$$

where $c_{gip}$ represents read-count depth at base pair $p$ within the gene $g$ in replicate $i$, $l_g$ is the length of gene $g$ and $\bar{c}_{gi} = \frac{1}{l_g} \sum_p c_{gip}$. When reads are randomly and uniformly distributed along the gene, $\chi^2_{gi} \sim 1$ is expected, but strongly non-uniform distributions give $\chi^2_{gi} \gg 1$. The median chi-square statistic calculated over all genes, $\tilde{\chi}^2_i$, gives a quantitative measure of the non-uniformity of a replicate and enables replicates to be compared on this basis within the same condition (Fig. 2c).

## Defining quality score

Interestingly, while there is similarity between the replicates identified as "bad" from the gene expression based quality measures (replicate correlation and outlier fraction, shown in Fig. 2, panels a and b) and those identified by examining the gene read depth profiles (Fig. 2c), there is no strict one-to-one correspondence between these three strongly related measures. Combining these metrics results in a simple replicate quality score that amplifies the distinctions between replicates considerably simplifying the identification of "bad" replicates. The score for each replicate is defined as:

$$S_i = A \log(1 - \tilde{r}_i) + B \log f_i + C \log(\tilde{\chi}^2_i - 1), \tag{2}$$

where $\tilde{r}_i$ is the median Pearson's correlation coefficient between the gene expression in replicate $i$ and the gene expression in all other replicates, $f_i$ is the fraction of genes identified as outliers in replicate $i$, and $\tilde{\chi}^2_i$ is the median reduced chi-square statistic for the gene read depth profiles in replicate $i$. For simplicity, the arbitrary weights of $A = B = C = 1$ are set here, though other prescriptions are possible. This quality score is shown in Fig. 2d for replicates of both conditions. Based on this score six WT replicates, and four Δ*snf2* replicates, are identified as "bad" (21, 22, 25, 28, 34, 36 and 6, 13, 25, 35, respectively). They were subsequently removed from the downstream analysis unless otherwise stated. In the following discussion, the remaining replicates are referred to as "clean replicates". This is an arbitrary and rather conservative selection. For example, one doesn't expect Δ*snf2* replicate 25 to make a huge impact. However, the most anomalous replicates in this selection affect the measured read-count distribution dramatically (see Section 3).

# 3 RESULTS

## 3.1 Inter-lane variance

The sample libraries were sequenced in seven of the eight lanes on an Illumina HiSeq2000, to give seven sequencing (technical) replicates for each biological replicate. This allowed testing of the inter-lane variance in the sequencing protocol. The RNA fragments from each sample were expected to be distributed uniformly and randomly across the flow cell lanes, which would result in a Poisson distribution of sequenced read counts across lanes (see Marioni et al., 2008 for a similar experiment) for a similar experiment). Any additional variability will result in a broader distribution.

Under the null hypothesis the read counts across lanes (for a given gene and biological replicate) are realizations of independent random variables that follow a Poisson law with the same mean (i.e., $\mu_{gi1} = \cdots = \mu_{gin_l}$). The appropriate statistic for the overdispersion test (Fisher, 1950) is chi-squared distributed. Applying this test to each gene in each replicate (all replicates were used for this test) with non-zero read counts in at least one lane resulted in a total of 630,850 tests, only two of which reject the null hypothesis with Holm-Bonferroni (Holm, 1979) corrected significance of $p < 0.05$. This confirms that the inter-lane sequencing variance is in excellent agreement with the expected uniform random distribution of sequenced RNA fragments across the lanes on the flow cell.

## 3.2 Inter-replicate variance

48 biological replicates in each of two conditions also allow the quantification of reproducibility in expression data across the replicates within a condition. Since the samples are independent, a Poisson distribution arises naturally from counting the number of reads mapping to each gene across replicates. Any additional variability, either from variability in the experimental protocol or biological variability, will result in broader distribution. The variance-mean relationship across all clean biological replicates (42 in WT and 44 in Δ*snf2*) is shown in Fig. 4. The data follow the Poisson law at lower mean count rates, but above ~10 counts per gene there is a smooth departure from the Poisson relationship and the data become over-dispersed. This can be represented by a dispersion parameter $\varphi = (\sigma^2 - \mu)/\mu^2$, where $\mu$ is the mean and $\sigma^2$ is the variance. The data approximately follow a constant dispersion, represented by the median dispersion parameter calculated over all genes (gold line). Though individual genes depart significantly from this line, the best-fitting Loess regression (blue line) agrees well with the constant dispersion parameter. This means that the relative width of the count distribution is the same across the full range of expression. The relative width of a probability distribution can be represented by its coefficient of variation, $\sigma/\mu$, and for large values of the mean $\varphi \approx (\sigma/\mu)^2$. In these data, the measured median dispersion coefficient is ~0.01, corresponding to a typical $\sigma/\mu \sim 0.1$.

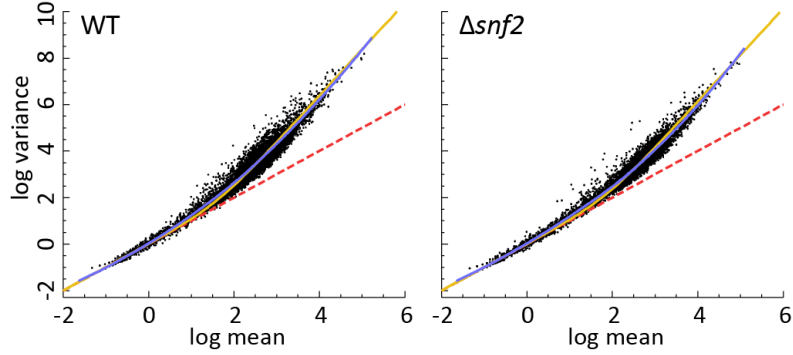

Fig. 4. Mean-variance relation for each condition. Each dot in the diagrams corresponds to one gene, with mean and variance calculated across all clean replicates in equal-count normalization. The red dashed line is the Poisson relation, where variance = mean. The gold curves show lines of constant dispersion parameter, $\phi$, calculated for the median dispersion in each condition: $\tilde{\phi}_{WT} = 2.35 \times 10^{-2}$ and $\tilde{\phi}_{\Delta snf2} = 1.35 \times 10^{-2}$. Blue curves show the smoothed data (in logarithmic space) using local polynomial regression fitting (R function 'loess' with second degree polynomials and smoothing parameter $\alpha = 0.75$).

### 3.3 Statistical models for read count data

Many differential expression tools address the over-dispersion inherent to short read RNA-seq data by assuming the form of the probability distribution underlying gene expression. Popular choices for this distribution include, but are not limited to, a negative binomial (NB, e.g. *DESeq* and *edgeR*) or log-normal (LN, e.g. *limma*) distribution. Here, the gene read counts from clean data are tested against log-normal, negative binomial and normal (NM) distributions.

The test of normality described by D'Agostino et al. (1990) was applied to examine whether the data are consistent with a normal distribution. It builds a statistic based on the skewness and kurtosis of the data and is sensitive to departures from normality. The same test was used to probe whether the gene expression data are consistent with a log-normal distribution after log-transforming the data. The disadvantage of this approach is that it cannot be applied to data containing zeroes, ruling it out for the ~10% of genes in each condition that contains at least one zero (766 out of 6873 genes in WT and 693 out of 6872 genes in Δ*snf2*). The expression data for all genes, including those with zeroes, was tested for consistency with a negative binomial distribution using the goodness-of-fit test proposed by Meintanis (2005). This test is based on the probability generating function and the distribution of the test statistic is not known in closed form so it requires a bootstrap to calculate *p*-values. Hence, it is more computationally intensive than the normality test. $10^7$ bootstraps were carried out and resulted in *p*-values that are limited to be $\gtrsim 10^{-7}$.

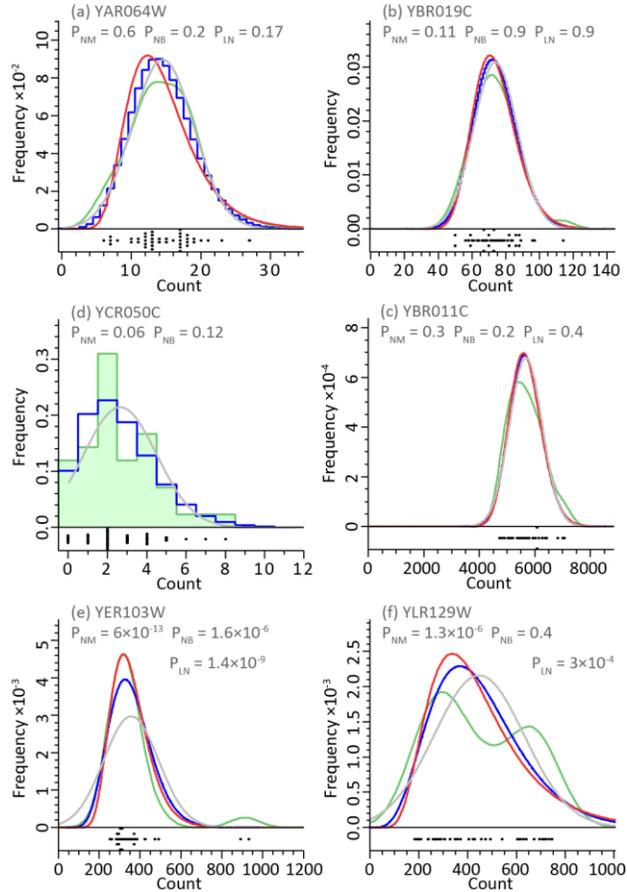

Fig. 5. Distribution of counts across biological replicates for six selected genes. All six examples are from the WT condition. The points represents equal-count normalized clean data. The green histogram and green lines show data histogram and kernel-density estimates, respectively. The grey, blue and red lines show best-fitting normal (NM), negative binomial (NB) and log-normal (LN) model distributions. The *p*-values shown are from respective goodness-of-fit tests, described in the text. (a-c) are the examples of low-, medium- and highly-expressed genes well fitted by all three distributions. (d) shows a low-count gene matched well by the NB distribution, but where no LN was possible due to zeroes in data. (e) is an example of data with strong outliers. These data are not consistent with NM, LN and NB distributions. (f) shows an unusual distribution of counts. It is strongly rejected by the NM and LN tests, but the NB test used is not sensitive to this type of distribution shape.

Fig. 5 shows six examples of gene read-count distributions across biological replicates. In each example, read-count data were fitted with all three models using the maximum likelihood method (Venables and Ripley, 2002) to illustrate the shape of their probability distributions. To quantify the level of agreement between data and each model the corresponding goodness-of-fit tests were performed and *p*-values calculated. The six examples shown include genes where the observed read-count distributions are consistent with all three models (Fig. 5a-c), an example of a gene that is consistent with the normal and negative binomial distributions, but cannot be tested against the log-normal distribution due to zeroes in some replicates (Fig. 5d), a gene with observed distribution that is formally inconsistent with all three models (Fig. 5e) and an unusual case of a bimodal observed distribution (Fig. 5f). Interestingly, due to the sensitivity of the d'Agostino test to small changes in the shape of the distribution, the example gene with a bimodal distribution is rejected by the log-normal and normal distribution tests with $p_{\text{LN}} = 3 \times 10^{-4}$ and $p_{\text{NM}} = 1.3 \times 10^{-6}$ respectively. The Meintanis negative binomial test, however, is insensitive to this type of distribution distortion and returns $p_{\text{NB}} = 0.4$. Note, that these *p*-values are for illustrative purposes only and are not corrected for multiple tests.

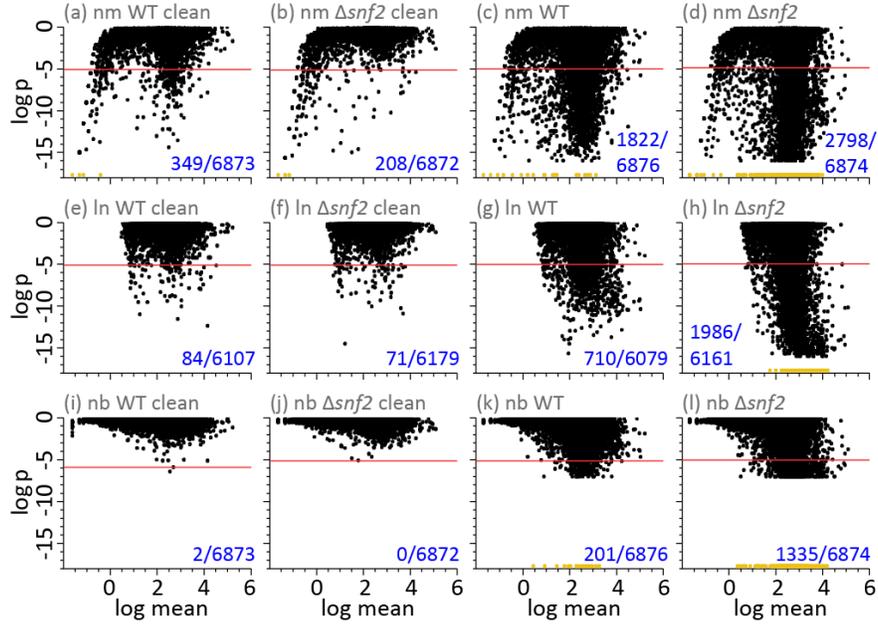

Fig. 6. Goodness-of-fit test results for normal (top panels), log-normal (middle panels) and negative binomial (bottom panels) distributions. Each panel shows the test *p*-value versus the mean count across replicates. Each dot represents equal-count normalized data from one gene. Panels on the left (a, b, e, f, i, j) show clean data with bad replicates rejected (42 and 44 replicates remaining in WT and Δ*snf2*, respectively). Panels on the right (c, d, g, h, k, l) show all available data (48 replicates in each condition). Due to the number of bootstraps performed, *p*-values for the negative-binomial test are limited to $\sim 10^{-7}$. Due to numerical precision of the software library used, *p*-values from the normal and log-normal tests are limited to $\sim 10^{-16}$. Below these limits data points are marked in orange at the bottom of each panel. Red horizontal lines show the Holm-Bonferroni limit corresponding to the significance of 0.05 for the given data set. The numbers in the right bottom corner of each panel indicate the number of genes with *p*-values below the significance limit and the total number of genes.

Fig. 6 illustrates the result of applying all the tests to the read-count data for all genes in both conditions. The null hypothesis is rejected using a Holm-Bonferroni corrected critical *p*-value of 0.05. The data are strongly consistent with both the log-normal and negative binomial distributions; only ∼ 1% of genes rejecting the log-normal hypothesis in either WT or Δ*snf2* data set, while the negative binomial test rejects this distribution in only 2 genes in WT (recalling, however, the lower sensitivity of the Meintanis negative binomial test). In fact, a large fraction of data is also consistent with a normal distribution, perhaps unsurprisingly given that both log-normal and negative binomial distributions approximate the normal distribution for high counts. At very low counts ($\bar{x}_g \lesssim 1$) data become increasingly non-normal due to its discrete nature, but remain consistent with the negative binomial distribution.

To assess the impact that "bad" replicates have on the observed distributions, the same calculations were also performed on the full set 48 replicates in each condition. When the "bad" replicates were included, the read counts for a large fraction of genes became inconsistent with all three model distributions. This demonstrates that bad replicates can have a profound distorting effect on read-count distribution and, by extension, differential expression results. The effect can be seen particularly clearly in the case of Δ*snf2*, where the fraction of genes inconsistent with a log-normal distribution increased from 1% to 41%, in comparison with clean data.

## 4  DISCUSSION AND CONCLUSIONS

The data presented here give us an unprecedented view of the variability inherent to RNA-seq experiments, albeit limited to an organism with gene expression largely unaffected by splicing and other complex mechanisms observed in higher eukaryotes. Distinguishing between variability introduced by experimental procedure and intrinsic biological variation is often tricky in RNA-seq experiments, and is complicated by the opaque use of the terms "technical replicate" and "biological replicate" in much of the literature. In the experiment presented here, 48 *S. cerevisiae* cell cultures were grown under the same experimental conditions for each of two biological conditions; the wild type (WT) and a *snf2* mutant line (Δ*snf2*). Each cell culture represents a distinct biological replicate of the given condition. However, each of the samples underwent a series of processing steps: RNA extraction, polyA selection, fragmentation, priming, cDNA synthesis, adapter ligation, PCR amplification and sequencing which can contribute to the overall variability. Accordingly, the observed read-count variability is a combination of true biological variability between individual samples, and the protocol variability introduced during sample preparation and sequencing.

### 4.1  Mean-variance relationship

In this work, the observed variance is tightly related to the mean and follows a line of constant dispersion parameter. Modelling the mean-variance relationship is essential in many DGE tools (e.g., in *edgeR* (Robinson et al., 2010) and *DESeq* (Anders and Huber, 2010)), when dealing with a small number of replicates. In such cases, variance at individual gene level has to be controlled by borrowing information from other genes. Our findings favour the approach implemented in *edgeR*, where variance for one gene is squeezed towards a common dispersion calculated across all genes.

### 4.2 Read-count distribution

A key finding of this work is the demonstration that the read-count distribution of the majority of genes is consistent with the negative binomial model. Reassuringly, many of the most widely used RNA-seq DGE tools (e.g., *edgeR*, *DESeq*, *cuffdiff*, (Trapnell et al., 2010) and *baySeq*, (Hardcastle and Kelly, 2010)) do assume a negative binomial distribution of gene-read count. Therefore, the majority of the RNA-seq DGE studies in the literature are, at least, based on tools that make appropriate assumptions. We recommend that future analyses of RNA-seq DGE experiments use existing, or new, tools that are based on a negative binomial distribution of gene read count. The log-normal model is also consistent with a large proportion of the genes in our data, however its use becomes problematic when one or more of the replicates contain zero counts.

### 4.3 Inter-lane variability

The design of the experiment presented here also allows us to probe the inter-lane sequencing variability in detail. This reflects how reads from one biological sample are distributed between different lanes of a flow cell and arises from a limited number of steps in the RNA-seq protocol, namely loading the samples into individual lanes, cluster amplification and sequencing by synthesis. An even loading and spatially uniform amplification and sequencing should give rise to a Poisson distribution of counts from individual genes. This is technical variability at a very low level. A $\chi^2$ dispersion test applied to inter-lane data confirms that they follow the Poisson law. This kind of test can be included in quality control to test for flow cell defects.

### 4.4 Impact of "bad" replicates

As shown in Fig. 6, one of the key observations from this study is the large impact that "bad" replicates have on the underlying properties of RNA-seq data. Without additional quality controls at the replicate level a large proportion of the data are inconsistent with all the statistical models tested. While the incidence of "bad" replicates can be reduced through improvements in sequencing techniques such as paired-end and longer reads, it cannot be entirely eradicated. In this study, 48 biological replicates were performed. This is not practical or cost-effective for routine experiments, but it is clear that increasing the number of biological replicates above the 2 or 3 typically performed in RNA-seq experiments will be beneficial in mitigating the risk of "bad" biological replicates skewing interpretation of the data.

**FUNDING**

This work was supported by: The Wellcome Trust (92530/Z/10/Z and strategic awards WT09230, WT083481 and WT097945), Biotechnology and Biological Sciences Research Council (BB/H002286/1 and BB/J00247X/1).

## ACKNOWLEDGEMENTS

We acknowledge Tom Walsh for his efforts and support in managing the software requirements for this project on our HPC platform. We also thank Simos Meintanis and Melanie Febrer for useful discussions.